# Using Quantum Confinement to Uniquely Identify Devices

J. Roberts[1], I. E. Bagci[2], M. A. M. Zawawi[3], J. Sexton[3], N. Hulbert[1], Y. J. Noori[1], M. P. Young[1], C. S. Woodhead[1], M. Missous[3], M. A. Migliorato[3], U. Roedig[2] and R. J. Young[1,*]

[1]Physics Department, Lancaster University, Lancaster, LA1 4YB, UK. [2]School of Computing and Communications, Lancaster University, Lancaster, LA1 4WA, UK. [3]School of Electrical and Electronic Engineering, University of Manchester, M13 9PL, UK.
*Correspondence and requests for materials should be addressed to R.J.Y. (r.j.young@lancaster.ac.uk).

Modern technology unintentionally provides resources that enable the trust of everyday interactions to be undermined. Some authentication schemes address this issue using devices that give unique outputs in response to a challenge. These signatures are generated by hard-to-predict physical responses derived from structural characteristics, which lend themselves to two different architectures, known as unique objects (UNOs) and physically unclonable functions (PUFs). The classical design of UNOs and PUFs limits their size and, in some cases, their security. Here we show that quantum confinement lends itself to the provision of unique identities at the nanoscale, by using fluctuations in tunnelling measurements through quantum wells in resonant tunnelling diodes (RTDs). This provides an uncomplicated measurement of identity without conventional resource limitations whilst providing robust security. The confined energy levels are highly sensitive to the specific nanostructure within each RTD, resulting in a distinct tunnelling spectrum for every device, as they contain a unique and unpredictable structure that is presently impossible to clone. This new class of authentication device operates with few resources in simple electronic structures above room temperature.

The rapid advance of manufacturing processes has provoked an accidental pathway to the creation of complex counterfeit components[1]. In tandem with this, there is an increasing menace from the processing power of modern computers, which can be utilised to mimic digital identities. Authenticating a device with a scheme such as certification[2] requires the use of a secret key acting as an identity, which is typically stored on an integrated circuit (IC). However, it has been shown that invasive and non-invasive attacks have the capability of learning this key, as it must exist in a digital form on the chip, and once compromised, an attacker can authenticate themselves as a legitimate device[3]. The IC can be protected by making it tamper-resistant, but this is expensive and difficult. Devices comprising randomness intrinsic to their fabrication can form the basis of solutions to the threat of hardware and software cloning, namely unique objects (UNOs)[4] and physically unclonable functions (PUFs)[5-7]. These devices form an inseparable link between their physical structure and an identity, providing a robust building block from which a secure system can be built. UNO's contain a unique fingerprint, with their security resting upon the impossibility of re-fabrication, with no restrictions on what an attacker may know about the internal structure or the fingerprint itself. Typical applications of UNOs include placing them on highly confidential documents such as bank notes, passports and access cards, where an attacker must be able to clone the subject to break their security. However, UNOs require a trusted external measurement apparatus every time fingerprint extraction is needed, which is undesirable.

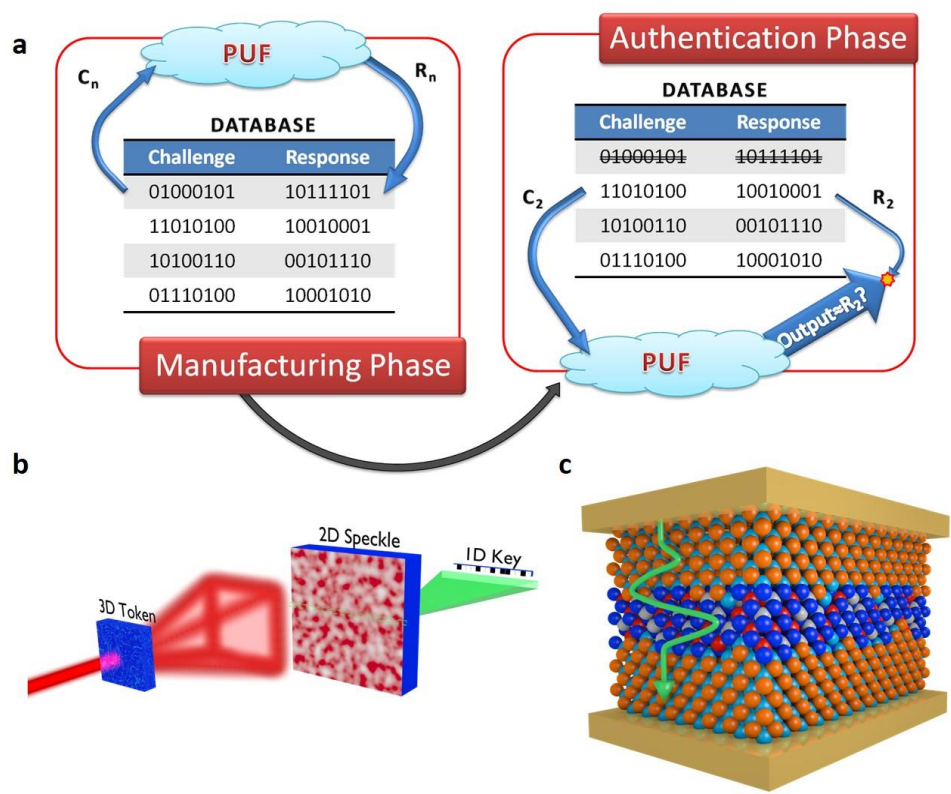

**Figure 1| Schematic, working principle and quantum analogue of a physically unclonable function (PUF). a,** An example operating protocol for a PUF. A database of challenge ($C_n$)-response ($R_n$) pairs is created by the manufacturer and stored online, the user can take a single entry from the database when required to check a device's authenticity. **b,** An optical PUF. The laser is dispersed by a three-dimensional object containing light scattering particles, this causes a two-dimensional speckled image to form, and this pattern can be transformed into a one-dimensional key using hash functions. **c,** Graphic of a conceptual UNO/PUF that relies on quantum-mechanical tunnelling through a quantum well containing imperfections (blue region).

PUFs are somewhat different, using disordered systems to derive a range of unique responses when challenged, which do not require digital storage. In addition to being used for low cost device authentication and identification, PUFs have several other uses, including secure key generation and binding software to hardware platforms. A method for using PUFs is demonstrated in Fig. 1a. A series of unique responses are generated by applying a variety of challenges to the PUF; these challenge-response pairs (CRPs) are used to authenticate the device[8]. Each CRP must be unique, unpredictable and repeatable whilst another device containing identical CRPs should be impossible to fabricate, even by the manufacturer. This approach requires a database where CRPs are recorded and used prior to each communication. Once used, a CRP is erased from the database; each pair being used just once. Moreover, a multiple CRP based authentication system requires a database that is large enough to meet security considerations. An alternative scheme, proposed by Koeberl et al., uses a single CRP to authenticate a device[9]. In this system the manufacturer stores a certificate which contains the sole response from the PUF within a signature signed with the manufacturer's private key. When authentication is required, the PUFs response is re-measured, whilst the signature is verified with the manufacturer's public key to extract the stored response. A check is then performed to determine whether the two values agree.

PUFs can be separated into two main types, weak PUFs (also known as physically obfuscated keys - POKs) and strong PUFs. Weak PUFs generate keys from a small set of CRPs, with the total set available scaling polynomially with size and complexity. In this architecture, the derived key normally remains secret through an internal measurement in the embedding hardware. In an ideal scenario the device is made tamper-proof to prevent knowledge of the stored key being determined by external and invasive attacks. Strong PUFs have a highly complex input-output behaviour, with the available set of CRPs scaling exponentially; their security relies upon an attacker not being able to determine this behaviour. On the contrary to weak PUFs, an entity is free to access a strong PUF and query its input-output behaviour whilst remaining unable to predict the response of a random challenge even if they have measured a large subset of CRPs. In both PUF systems, the CRPs should be stable under repeated measurements and changing environmental conditions. A number of methods have been proposed to construct both UNOs and PUFs, including; scattering from an optical medium (illustrated in Fig. 1b)[5], modes in silicon ring oscillators[6], statistical delay variations between nominally identical paths[10,11] and the state of static random access memories (SRAM) cells[12,13]. However, some constructions are vulnerable to simulation and cloning amongst other attacks. For example, an SRAM PUF was successfully cloned within a period of 20 hours by Helfmeier *et al.*[14], arbiter PUFs and their evolutions have been shown to be susceptible to machine learning[15] and a number of other PUFs have demonstrated vulnerabilities to side-channel attacks[16].

As the size of a system reduces, a limit is reached at which quantum confinement starts to govern the properties of the system and here the nanostructure of the atomic layers can become crucial to its properties[17]. As the confined energy levels are extremely sensitive to these layers that contain millions of atoms, the probability of creating a unique device is extremely high due to the inherently random nature of the atomic positions and imperfections, as illustrated in the quantum well in Fig. 1c. Simulating these structures requires vast computing power and is not achievable on a reasonable timescale, even with a modest quantum computer[18, 19]. When coupled with the fact that the underlying structure is unknown, unless dismantled atom-by-atom, this makes simulation extremely difficult. Given the impracticality of copying the device at the atomic level, such technology would provide near guaranteed unclonability. A quantum well represents the 'least-unique' quantum structure, with one dimension of confinement, but it enables us to demonstrate the proof-of-principle. The application of quantum phenomena in UNO/PUF-like architectures provides a means of harbouring a secret identity on the nanoscale in devices that can be incorporated in current microelectronic processes. This enables simple system integration whilst having lower size, weight and power footprints than current systems.

To realise a quantum mechanical UNO/PUF we use a simple device that can measure phenomenological properties arising from quantum confinement. The implementation of quantum tunnelling can be readily achieved by using a resonant tunnelling diode (RTD) containing a quantum well. These are double-barrier structures that allow electrons to tunnel through directly at voltages where the energy level within the quantum well lines up with the conduction band minimum. The confined energy level is exponentially sensitive to the width and height of the well and the barriers and as such on the atomic uniformity predominantly at the interfaces between layers[20, 21, 22, 24]. A measurement to find currents corresponding to these energies can be made by sweeping the voltage through a range. As shown in Fig. 2c, the Stark shift that results from the application of a voltage across the diode causes the energy levels within the well to lower, moving into resonance with the conduction band minimum and resulting in a peak in current. This subsequently diminishes as the bias is increased. The resultant room-temperature spectrum from a typical device is shown in Fig. 2b.

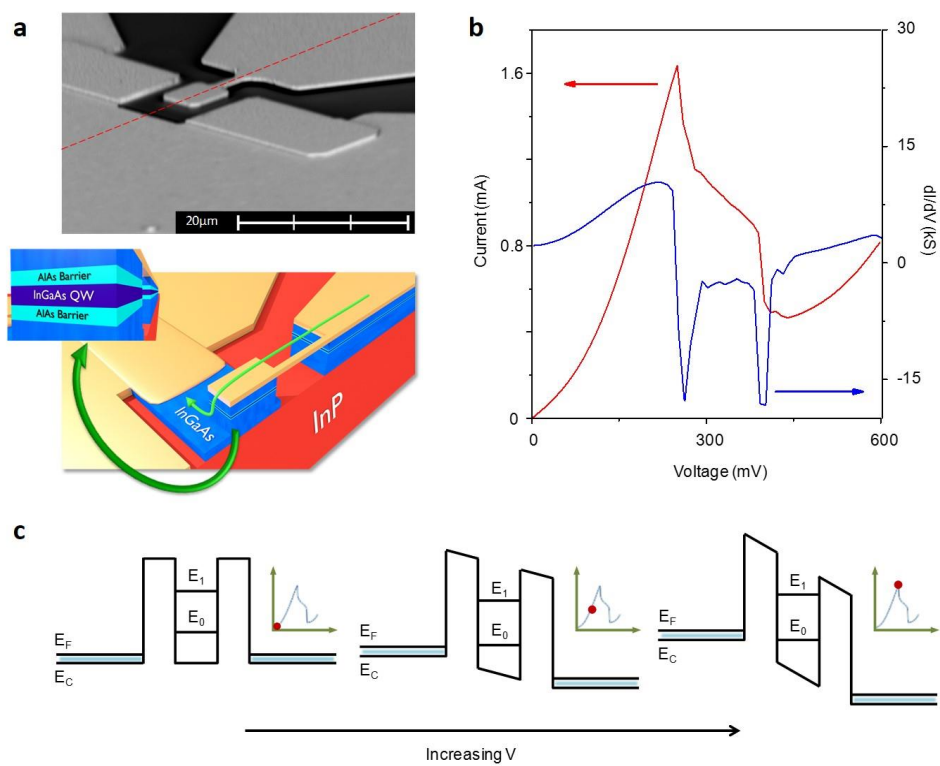

**Figure 2| Structure, I-V characteristic and band diagram of a resonant tunnelling diode (RTD). a,** Scanning electron microscopy image of a typical device (top) and a rendered counterpart of the cross-section through the red dashed line (bottom) with an inset showing the active region to highlight the important features of the sample; an InGaAs quantum well and barriers made of AlAs **b,** A representative I-V (red) and dI/dV (blue) spectrum from an RTD; the peak in current arises due to the resonance of the confined energy level with the conduction band minimum of the system **c,** Schematic of the E-k structure of the quantum well as the voltage is increased, demonstrating the nature of resonant tunnelling.

Typical tunnelling devices exhibit variations of 5% or more in their I-V characteristics[23]. However, considerable effort has been made to produce highly uniform I-V spectra from tunnel barriers. State-of-the-art manufactured devices have only a 0.02 monolayer variation over an 8" wafer and result in a variation of less than 1% in their I-V characteristics (about a pre-specified mean)[24]. However, the devices presented in that work used the most commonly studied single barrier binary-binary structure and thus only rely on the order of two interfaces. The RTDs we present here use 2 barriers so depends on four interfaces whilst also incorporating a ternary material into the quantum well. The result is a much larger interfacial roughness at the two binary-ternary interfaces resulting in fluctuations in the position and width of the confined energy level from device–to-device[20,22,25].

In this work we explore the distinct I-V characteristics that arise due to the sensitivity of the confined energy levels within quantum wells contained in resonant tunnelling diodes (RTDs)[18, 26, 27] and show they can provide robust measurements for unique device applications without typical size restraints. Furthermore, the nanostructure within an RTD is impossible to clone with current techniques[28,29].

**Results**

For the measurements presented here, RTDs fabricated with mesa sizes of 2x2 µm² were studied - the small area for this geometry of device was found to improve stability. An illustration of the structure used is shown in Fig. 2a, alongside an SEM micrograph; the gold region in the bottom-left represents the back contact whereas the RTD containing mesa is connected via an air bridge to the top contact in the top-right of the images. The important aspects of the structure lie in the region where resonant tunnelling takes place; an InGaAs quantum well between two AlAs barriers, shown in the inset to Fig. 2a. To justify quantum tunnelling as a measure of uniqueness, 26 devices were manufactured and tested with nominally identical features, namely a rounded mesa connected to a 3µm air bridge. Each device's average current-voltage characteristic was measured and analysed with a Gaussian fit to find the peak current and voltages, as plotted in Fig. 3a. From this figure we note that there is a large scope of available peak positions, ranging by approximately 70 mV in voltage and 4 mA in current. The exact position of devices within this region is unique and the apparent overlap in the highlighted area is an artefact of the symbol size. Fig.'s 3c and 3d investigate this area; the former shows the true precision of the calculated average using ellipses to represent confidence bounds. These were extracted from the spread of measurements. Upon examination of Fig. 3c there is no overlap between devices with 99.997% certainty. The precision of the average to these confidence ellipses was found by using the standard errors in voltage as the semi-major axis and standard errors in current as semi-minor axis. The ellipses confidence limits of 95%, 99.9% and 99.997% correspond to using 1.96, 3.09 and 3.99 standard errors respectively. The variation in the height and width of these ellipses is a direct result of the varying interfacial roughness from device to device. These imperfect interfaces result in the energy level within the well broadening on a different scale for each device, and thus they lead to a device-dependent scattering in the current and voltage measurements.

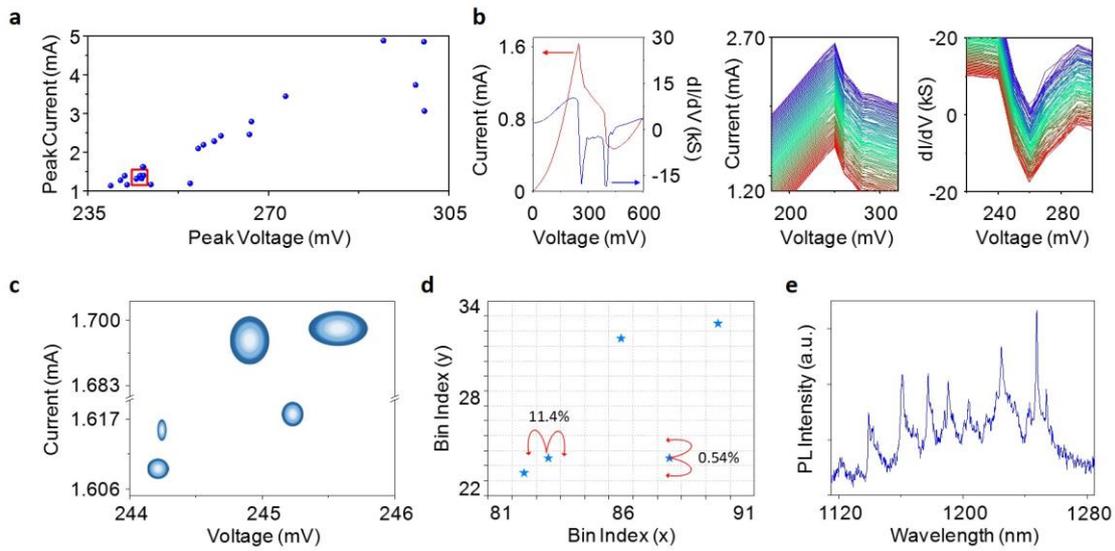

**Figure 3| Uniqueness and reproducibility performance of an RTD PUF. a**, Positions of associated peak voltages and currents for 26 devices manufactured to have identical characteristics. **b**, I-V and dI/dV curve of a single device (left); 100 measurements of I-V (centre) and dI/dV (right) from the same device (offset for clarity). **c**, Zoomed-in view of the highlighted section in **a** showing 95% (inner), 99.9% (middle) and 99.997% (outer) confidence ellipses of the 5 devices tested that lie the closest together. **d**, Probability of a device falling into another bin on x/y axis for the same area as in **c**. **e**, Photoluminescence spectrum from a sample containing quantum dots.

The motivation behind re-plotting the data from Fig. 3c on axes represented by bin indices in Fig. 3d lies in the realistic implementation of a device. In practical execution, a unique number would be extracted depending on bin position. We have split both axes into 256 bins and considered the probability of a device changing its bin index when re-measured; this is 11.4% on the x-axis and 0.54% on the y-axis. These probabilities are largely dependent on the number of bins on a particular axis, with more bins the probability of shifting increases, but the total number of potential devices also increases.

Measurements need to be reproducible when considering implementation; the results shown in Fig. 3b were taken to test the stability of these devices. The left figure is the average I-V characteristic for an RTD and from this the peak in both the current (centre) and differential current, or gradient of the negative differential resistance region (right) vs. voltage is displayed in more detail. The two graphs show data from 100 repeated measurements of the device. The peak position of each measurement has been found to lie within two standard errors of the calculated average in both measurement axes, an indication of the high calibre of robustness expected from such devices.

**Discussion**

The RTD structures introduced in this work represent uniqueness on the quantum scale, providing unclonability with the smallest size, weight and power requirements. Although the mesoscopic properties of the devices certainly play a role (metal layer thickness, device geometry, contact interfaces etc), this does not detract from the fact that on the nanoscale, due to the disordered nature of the binary-ternary interfaces, each device is unique. The mesoscopic properties only add to this uniqueness, though these properties can be adjusted and controlled much more stringently. This is evidenced in ref. 22 where they reduced the variation in the I-V characteristics to less than 1% by solely refining the interfaces between layers and leaving the fabrication process unchanged. If one wanted to make this type of device even more unique, one could manipulate the MBE growth, by not rotating the sample stage during a certain stage for example. This will result in interfacial variations of the quantum well or barrier on the order of 1 monolayer, corresponding to changes in the I-V characteristics of approximately 270%[24]. Moreover, this system would display effective tamper-resistance as any effort to invade the device would largely distort the nanostructure and hence the produced results. However, this 'nano-rearrangement' could also be used by an honest party to 'reset' the device, by a controlled application of heat, for example; as in reconfigurable PUFs[30]. A requirement for PUF architectures is the simplicity of extracting the secret information harboured within the device. Devices must be both straightforward to manufacture and utilise an easily obtainable measurement. We anticipate that both these features are readily achievable with the suggested RTD based device. The complexity and scalability of state-of-the-art epitaxy and lithography techniques allows thousands of high-quality but unique devices to be fabricated with an identical process. Moreover, conventional CMOS circuitry can be adapted to integrate the devices into embedded systems that can evaluate the internal structure without difficulty. In its current form, the RTD based system is an example of a weak PUF, due to the linear growth in the number of CRPs as the number (or area) of devices is scaled. To create a strong PUF, a challenge-response database with exponential growth is needed. This could be achieved by coupling a number of RTDs together in series[31-33], the challenge would correspond to which RTDs are linked and the response would be a multi-peak I-V spectrum that depends on the randomly linked RTDs. The number of CRPs should then increase exponentially with the number of devices that are coupled. A previous example of this idea is shown in the work by Rührmair *et al.* in which they suggest using crossbar structures to link a series of conventional diodes[34]. Another interesting concept is to exploit changes in the PUF response due to the local variations in temperature and magnetic field, as they are not known to an attacker. This would allow the production of a secret key in an offline system, in which a database check is not performed, even if an identical challenge was used, simply due to these local fluctuations.

Taking the current range spanned by the devices measured here, and the average uncertainty in the peak position, measured with high confidence, we can extrapolate that the PUF structure introduced here could provide around $10^3$ unique identities. For practical applications, such a number would be easily increased, by combining multiple devices in an array.

The use of three-dimensional nanostructures would also significantly increase the uniqueness of a device. As an example of such nanostructures, quantum dots typically have many electron and hole confinement levels, as illustrated by the rich photoluminescence (PL) spectrum emitted from GaSb quantum dots shown in Fig. 3e. RTDs containing single, or a few, dots reflect this increase in the number of confined states with an increased number of peaks in their dI/dV curves[35-37]. Each peak, when fitted individually and combined, could form part of a unique key for the device. The benefit of applying quantum dots within a resonant tunnelling structure is the practicality of such an electronic room-temperature measurement (a necessity for PUFs). Moreover, as the number of dots in a device does not need to be reproducible, fabrication using self-assembly techniques are well suited here.

It could be argued that the devices presented here do not necessarily need to be cloned, and all that is needed is a similar device that can be modified, for example by a gated electric field, to produce a similar I-V peak position. However, this is only the case if we naïvely look at a single peak in the I-V characteristic, as in this proof-of-principle demonstration. In a true implementation there are many other degrees of freedom, including the position of the valley, the gradient of the NDR region and the FWHM of the peak. Furthermore, the device will exhibit an asymmetric peak in which different values of the above can also be measured in reverse bias. If we assume all of the above can still be individually varied to replicate another device, then we can complicate the system further by coupling a number of these devices together (whilst incorporating quantum dots in the active region) as mentioned earlier in the discussion. Furthermore, if the application of the device is a UNO then this kind of attack is impossible, because the measurement device must be trusted and it could (visually) be checked that the device has not been tampered with externally (such as attaching a gate).

While inhomogeneity in the fabrication of nanostructures often leads to unpredictable behaviour of the final device, which is normally undesirable, we have proposed and demonstrated a potential use for the quantum behaviour of atomically irreproducible systems. The devices presented, based around 1D quantum structures, afford a secure bit density of 2.5 bits/$\mu m^2$. This is twice the value of state of the art classical PUFs[38], and will increase significantly for devices containing structures that provide three-dimensional quantum confinement. These devices can be seamlessly integrated into embedded electronic systems to provide robust unique identities requiring atom-level engineering to clone.

It should be noted that a security analysis of our PUF architecture has not been shown here for a number of reasons. One being that this letter first and foremost details a proof-of-principle of a new type of PUF that has a much lower overhead than existing architectures. Furthermore, the number of devices that were studied here was limited to 26; this being the amount that were 'identical' on chip. This is not enough for a formal statistical analysis of the security of this PUF system. We also realise that error-correction techniques would need to be applied to ensure the same output is achieved from each device, this has also been omitted here and will be included with the security analysis in future work.

**Methods**

The RTD devices were fabricated from an InGaAs/AlAs double-barrier structure grown by molecular beam epitaxy on an InP substrate. The details of this are given in ref. 39. To fabricate the RTDs, a top contact was first defined using conventional i-line optical lithography. A non-alloyed ohmic contact method was employed, where titanium (50 nm) and gold (250 nm) were deposited onto the surface of the highly doped cap layer by thermal evaporation. The top metal itself acted as a hard mask for a subsequent mesa etch. A reactive-ion etching (RIE) process using a mixture of methane ($CH_4$) and hydrogen ($H_2$), with an etch rate of 21 nm/minute, was implemented in order to produce anisotropic side-walls to the bottom contact layer, in preparation for the bottom metal contact deposition. Just before the bottom metal contact process took place, a non-selective orthophosphoric-based ($H_2O:H_3PO_4:H_2O_2$ = 50:3:1) wet-etch with a etch rate of 90 nm/minute was used to etch away 200 nm of epilayers down to the surface of the InP to completely isolate neighbouring devices. The 5 minutes wet-etch also simultaneously provided the necessary undercut for the air-bridge formation. Finally, the bottom ohmic contact was formed by thermal evaporation of Ti/Au (50 nm/500 nm).

The devices measured were not precisely 2x2µm², as the fabrication process tends to resulted in round-shaped mesas, however this is a minor detail as the measurements made are much more sensitive to the variations in the 1D confinement potential of the well than its profile. All measurements were taken at 300K using a Keithley 2400 source measure unit connected to a Wentworth Laboratories Ltd. SPM197 probe station using two 1.25" tungsten probes with 1µm tip radii. For each RTD, a voltage sweep between 0 and 1V was performed with the current being measured in voltage steps of 10mV; this measurement was repeated 100 times per device. The voltage sweep was performed as follows: the voltage source is initially set to a value of 0V, a measurement delay of 80ms is used to allow the source to settle to the given source value and subsequently the average current measurement corresponding to this voltage was taken over 60ms, finally the source voltage is stepped up by 10mV and this whole process is repeated for the next value. Because of the nature of two probe measurements, the evaluated current-voltage characteristics showed oscillations in the peak values due to the probes making intermittent contact with the surface (causing changes in resistance), therefore the I-V curves were re-taken until a good contact was achieved. The voltage and current ranges were also key considerations; if probed above 1V or allowed to rise above 10mA the RTDs broke down and then showed uncharacteristic ohmic-like behaviour. Finally, it is important to note that during 100 measurements, it was clear that some devices were reaching a critical temperature, which resulted in small chemical changes that seemed to cause a slight shift in the peak I-V. This particular aspect could be useful when considering a realistic implementation, as it would enable the device to be effectively 'reset' so that it exhibited a distinct new signal.

# References

[1] Guin, G. *et al*. Counterfeit integrated circuits: a rising threat in the global semiconductor supply chain. *in Proc. IEEE* **102**, 1207-1228 (2014).

[2] Simmons, G. J. *Contemporary Cryptology: The Science of Information Integrity* (IEEE Press, New Jersey, 1994).

[3] Barenghi, A., Breveglieri, L., Koren, I. & Naccache, D. Fault injection attacks on cryptographic devices: theory, practise and countermeasures. in *Proc. IEEE* **100**, 3056-3076 (2012).

[4] Rührmair, U., Devadas S., Koushanfar F. Security Based on Physical Unclonability and Disorder. In: Introduction to Hardware Security and Trust, 65-102 (Springer, New York Dordrecht Heidelberg London, 2011)

[5] Pappu, R., Recht, B., Taylor, J. & Gershenfeld, N. Physical one-way functions. *Science* **297**, 2026-2030 (2002).

[6] Gassend, B., Clarke, D., van Dijk, M. & Devadas, S. Silicon physical random functions. *Paper presented at Computer and Communication Security Conference, Washington DC, USA.* DOI: 10.1145/586110.586132 (2002, Nov. 18-22).

[7] Tehranipoor, M. & Wang, C. *Introduction to Hardware Security and Trust*, 67-74 (Springer, New York Dordrecht Heidelberg London, 2012).

[8] Suh, G. E. & Devadas, S. Physical unclonable functions for device authentication and secret key generation. *Paper presented at Design Automation Conference, San Diego, CA, USA.* DOI: 10.1145/1278480.1278484 (2007, June 4-8).

[9] Koeberl, P., Li, J., Rajan, A. Vishik, C. & Wu, W. A practical device authentication scheme using SRAM PUFs. in *Proc. of the 4th International Conference on Trust and Trustworthy Computing,* 63-77 (Springer 2007).

[10] Lee, J. W. *et al.* A technique to build a secret key in integrated circuits for identification and authentication application. *IEEE Symposium on VLSI Circuits,* 176-179 (IEEE 2004).

[11] Lim, D. *et al.* Extracting secret keys from integrated circuits. *IEEE Transactions on VLSI Systems* **13**, 1200-1205 (2005).

[12] Guajardo, J., Kumar, S. S., Schrijen G. J. & Tuyls, P. FPGA intrinsic PUFs and their use for IP protection. *Paper presented at Cryptographic Hardware and Embedded Systems, Vienna, Austria.* Berlin: Springer (2007, Sept. 10-13).

[13] Holcomb, D. E., Burleson, W. P. & Fu, K. Power-up SRAM state as an identifying fingerprint and source of true random numbers. *IEEE Transactions on Computers* **58**, 1198-1210 (2009).

[14] Helfmeier, C., Nedospasov, D., Boit, C. & Seifert, J. Cloning physically unclonable functions. *Paper presented at IEEE International Symposium on Hardware-Oriented Security and Trust, Austin, TX, USA.* DOI: 10.1109/HST.2013.6581556 (2013, June 2-3).

[15] Rürrmair, U. *et al.* Modeling attacks on physical unclonable functions. *Paper presented at Conference on Computer and Communications Security, Chicago, IL, USA.* DOI: 10.1145/1866307.1866335 (2010, Oct. 4-8).

[16] Merli, D., Schuster, D., Stumpf, F. & Sidl, G. Side Channel Analysis of PUFs and Fuzzy Extractors. *Paper presented at 4th International Conference on Trust and Trustworthy Computing, Pittsburg, PA, USA.* DOI: 10.1007/978-3-642-21599-5_3 (2011, June 22-24).

[17] Tsuchiya, M. & Sakaki, H. Dependence of resonant tunneling current on well widths in AlAs/GaAs/AlAs double barrier diode structures. Appl. Phys. Lett. 49, 88 (1986).

[18] Cirac, J. I. & Zoller, P. Goals and opportunities in quantum simulation. *Nature Phys.* **8**, 264-266 (2012).

[19] Buluta, I. & Nori, F. Quantum Simulators. *Science* **236**, 108-111 (2009).

[20] Wilkinson, V. A., Kelly, M. J. & Carr, M. Tunnel devices are not yet manufacturable. *Semicond. Sci. Technol.* **12**, 91-99 (1997).

[21] Kelly, M. J. New statistical analysis of tunnel diode barriers. *Semicond. Sci. Technol.* **15**, 79-83 (2000).

[22] Dasmahapatra, P., Sexton, J., Missous, M., Shao, C. & Kelly, M. J. Thickness control of molecular beam epitaxy grown layers at the 0.01-0.1 monolayer level. *Semicond. Sci. Technol.* **27**, 085007 (2012).

[23] Shao, C., Sexton, J., Missous, M. & Kelly, M. J. Achieving reproducibility needed for manufacturing semiconductor tunnel devices. *Electronics Letters* **49**, 10 (2013).

[24] Missous, M., Kelly, M. J. & Sexton, J. Extremely uniform tunnel barriers for low-cost device manufacture. *IEEE Electron Device Letters* **36**, 6 (2015).

[25] Kelly, M. J. The unacceptable variability in tunnel current for proposed electronic device applications. *Semicond. Sci. Technol.* **21**, L49-L51 (2006).

[26] Lin, Y., van Rheenen, A. D. & Chou, S. Y. Current fluctuations in double-barrier quantum well resonant tunneling diodes. *Appl. Phys. Lett.* **59**, 1105 (1991).

[27] Ng, S., Surya, C., Brown, E. R. & Maki, P. A. Observation of random-telegraph noise in resonant-tunneling diodes. *Appl. Phys. Lett.* **62**, 2262 (1993).

[28] Wegscheider, W., Schedelbeck, G., Abstreiter, G., Rother, M. & Bichler, M. Atomically precise GaAs/AlGaAs quantum dots fabricated by twofold cleaved edge overgrowth. *Phys. Rev. Lett.* **79**, 1917-1920 (1997).

[29] Fölsch, S., Martinez-Blanco, J., Yang, J., Kanisawa, K. & Erwin, S. C. Quantum dots with single-atom precision. *Nature Nanotech.* **9**, 505-508 (2014).

[30] Kursawe, K., Sadeghi, A., Schellekens, D., Škorić, B. & Tuyls, P. Reconfigurable physical unclonable functions – enabling technology for tamper-resistant storage. *Paper presented at IEEE International Workshop on Hardware-Oriented Security and Trust, San Francisco, CA, USA*. DOI: 10.1109/HST.2009.5225058 (2009, July 27).

[31] Sen, S., Capasso, F., Cho, A. Y. & Sivco, D. Resonant tunneling device with multiple negative differential resistance: digital and signal processing applications with reduced circuit complexity. *IEEE Transactions on Electron Devices*, **34**, 10 (1987).

[32] Yan, Z. X., Fraser, S. & Deen, M. J. A new resonant-tunnel diode-based multivalued memory circuit using a MESFET depletion load. *IEEE Journal of Solid-State Circuits*, **27**, 8 (1992).

[33] Waho, T., Chen, K. J. & Yamamoto, M. A novel multiple-valued logic gate using resonant tunneling devices. *IEEE Electron Device Letters*, **17**, 5 (1996).

[34] Rürrmair, U. *et al.* Security applications of diodes with unique current-voltage characteristics. *Lecture Notes in Computer Science,* **6052**, 328-335 (2010).

[35] Li, P. W., Kuo, D. M. T. & Hsu, Y. C. Photoexcitation effects on charge transport of Ge quantum-dot resonant tunneling diodes. *Appl. Phys. Lett.* **89**, 133105 (2006).

[36] Lai, W., Kuo, D. M. T. & Li, P. Transient current through a single germanium quantum dot. *Physica E* **41**, 886-889 (2009).

[37] Chen, K., Chien, C. & Li, P. Precise Ge quantum dot placement for quantum tunneling devices. *Nanotechnol.* **21**, 055302 (2010).

[38] Maes, R. *Physically Unclonable Functions* (Springer, Berlin, 2013).

[39] Zawawi, M. A. M., KaWa, I., Sexton, J. & Missous, M. Fabrication of sub-micrometer InGaAs/AlAs resonant tunneling diode using a tri-layer soft reflow technique with excellent scalability. *IEEE Transactions on Electron Devices*. **61**, 2338-2342 (2014).

**Acknowledgements** This work is supported by the Royal Society through a University Research Fellowship held by R.J.Y. J.R. is supported by the EPSRC 'NOWNANO' DTC. M.M. also acknowledges the support of the Science and Technologies Facilities Council (STFC). R. J. Y. would like to acknowledge useful discussions with Ulrich Rührmair of Technische Universität München, Germany.

**Author Contributions** R.J.Y. and U.R. designed the project. M.M., J.S and M.A.M.Z. fabricated and performed initial tests on the RTD devices. J.R, N.H., C.S.W., Y.J.N. and M.P.Y. performed the detailed electronic measurements. J.R., I.E.B., U.R. and R.J.Y. analysed the data. M.A.M. provided theoretical support. The manuscript was prepared primarily by J.R. and R.J.Y., with contribution from all authors.

**Competing Financial Interests** There is no competing financial interests for this work.